\documentclass[12pt]{article}
\usepackage{graphicx,epsf}
\usepackage{float}
\newcommand{\y }{\'{\i}}

\begin{document}

\title {Surface scaling analysis of hydrogels: From multiaffine 
to self-affine scaling}
\author{
G.M.~Buend{\y}a$^1$,
S.J.~Mitchell$^2$,
P.A.~Rikvold$^3$\\
$^1$Departamento de F{\y}sica,
Universidad Sim\'on Bol{\y}var,\\
Caracas 1080, Venezuela\\
$^2$Shuit Institute for Catalysis
and Department of Chemistry \\
and Chemical Engineering,
Eindhoven University of Technology,\\
5600 MB Eindhoven, The Netherlands\\
$^3$Center for Materials Research and Technology, \\
School of Computational Science and Information Technology,\\
and Department of Physics,
Florida State University, \\
Tallahassee, Florida 32306-4350, USA\\
}
\maketitle

\begin{abstract}
We show that smoothing of multiaffine surfaces that are 
generated by simulating a crosslinked polymer gel by a frustrated,
triangular network of springs of random
equilibrium lengths [G.M.\ Buend{\'\i}a, S.J.\ Mitchell, P.A.\ Rikvold, 
Phys.\ Rev. E 66 (2002) 046119] changes the scaling behavior of the
surfaces such that they become self-affine. The 
self-affine behavior is consistent with
recent atomic force microscopy (AFM) studies of the surface
structure of crosslinked polymer gels into which voids are
introduced through templating by surfactant micelles 
[M.~Chakrapani, S.J.\ Mitchell, D.H.\ Van Winkle, P.A.\ Rikvold, 
J.\ Colloid Interface Sci., in press].
The smoothing process mimics the effect of the AFM tip
that tends to flatten the soft gel surfaces. Both the
experimental and the simulated surfaces have a non-trivial
scaling behavior on small length scales, with a crossover to
scale-independent behavior on large scales.
\end {abstract}



\section{Introduction}

In a recent study \cite{GSP}
we proposed a frustrated spring-network model to simulate the surface structures
of crosslinked polymer gels into which voids are introduced through 
templating by surfactant micelles.
Experimentally, templated polyacrylamide gels are created by mixing acrylamide
with a crosslinker in the presence of a surfactant \cite{exp1}.
The resultant gels have a wide range of pore sizes,
making them ideal materials for macromolecular separation \cite{general}.
The crosslinking process introduces inhomogeneities in the spatial density
of the gel that affect its surface configuration.
Our results show that the surfaces generated by the spring-network model
have a nontrivial, multiaffine scaling behavior at small length scales, 
with a crossover to scale-independent behavior on large scales.

At small length scales we found that the $q$th order increment
correlation function has a power-law behavior, with $q$-dependent
exponents that change with the vacancy concentration. However,
experimental surfaces of surfactant-templated hydrogels studied by
atomic force microscopy  (AFM) \cite{exp2} show $q$-independent 
self-affine scaling at length scales below the
crossover length. In Ref.~\cite{GSP} we discussed the possibility that
the multiaffine behavior was due to a finite density of vertical
discontinuities due to overhangs in the simulated surfaces. Later work showed
analytically and numerically that when vertical discontinuities
are introduced into a self-affine surface, it becomes
multiaffine \cite{Steven}. Here we extend this
study to hydrogel surfaces, and we show that smoothing the simulated 
surfaces by convoluting them with a Gaussian changes
the scaling behavior from multiaffine to self-affine, in agreement with the
experimental AFM results. The smoothing process mimics the effect of
the measurement process of an AFM tip interacting with the soft
substrate.

\section{The frustrated spring-network model}

The model, described in detail in Ref.~\cite{GSP}, consists of a
two-dimensional triangular network of nodes interconnected by
massless harmonic springs. The nodes represent the crosslinkers,
and the springs correspond to the polymer chains. We impose
periodic boundary conditions in the horizontal direction, while the top
nodes are free to move in two dimensions. The bottom nodes are
fixed, corresponding to bonding to a rigid substrate.

We assume that each spring corresponds to a collapsed polymer
chain with equilibrium end-to-end length $l_{0i}$, which is randomly chosen with
probability density
\begin{equation}
P(l_{0i})=2\gamma l_{0i} \exp(-\gamma l_{0i}^{2}) 
\end{equation}
Thus, $l_{0i}$ is proportional to the square root of an exponentially distributed
number of monomers between crosslinkers \cite{pol1}.
The constant $\gamma$ is proportional to the inverse of the
average number of monomers between crosslinkers. Consistent
with the picture of a random-coil collapsed polymer chain, we
require that the elastic constant of each spring, $k_{i}$, be
proportional to its equilibrium length, $k_{i}=1/l_{0i}$ in
dimensionless units \cite{pol2}.

For each statistical realization of the system,
the equilibrium spring lengths are chosen independently according to Eq.~(1).
In the initial network configuration,
all springs are placed on the bonds of a regular triangular lattice of unit lattice constant,
such that all springs are stretched or compressed to unit length, $l_{i}=1$.
This initial configuration is highly stressed.
However, by choosing the value of $\gamma$
such that the average force exerted by each spring is zero,
we ensure that the initial configuration is not {\it globally\/} stressed.
{}From the initial configuration,
the system is relaxed to one of its many local energy minima by a
limited-memory Broyden-Fletcher-Goldfarb-Shanno (L-BFGS)
quasi-Newton minimization algorithm \cite{LM},
and by repeating the calculations for several different realizations
of the equilibrium spring lengths we verified that the statistical 
properties of the different relaxed configurations are essentially the same.

Spring networks of $L_{x}\times L_{y}=1024\times 768$ nodes were
generated, and voids were introduced by randomly removing nodes
and the springs connected to them. The Hoshen-Kopelman algorithm \cite{HK}
was applied to eliminate clusters that were not connected 
to the fixed substrate, thus mimicking the removal of non-bonded
material which is washed away during the experimental preparation.

\section{Calculations}

The one-dimensional surfaces are defined as the set of surface
heights, $h(x_{i})$, at equally spaced discrete horizontal points,
$x_{i}$. The surface height at $x_{i}$, $h(x_{i})$, is
taken as the largest vertical distance, measured from the bottom,
of the intersections of the springs with a vertical line at
$x_{i}$, where each spring is represented by a straight line segment  
between the connected nodes. This definition of the surface is
described in more detail in Ref.~\cite{GSP}.

The $q$th order 
generalized increment correlation function is defined as \cite{self1}
\begin{equation}
C_{q}(r)=\langle | h(x_{0}+r)-h(x_{0})| ^{q}\rangle 
\end{equation}
For many surfaces, $C_q$ displays a power-law behavior, such that
\begin{equation}
C_{q}(r)\propto r^{qH_q} 
\end{equation}
where $H_{q}$ is the generalized Hurst exponent \cite{self1,self2}. For
self-affine surfaces, the scaling behavior is independent of $q$,
$H_{q}=H$, but for multiaffine surfaces, there is an infinite
hierarchy of scaling exponents, such that $H_{q}$ depends
continuously on $q$, at least for some range of $q$ values and
length scales \cite{multi1}.

Very often surfaces present a mixed behavior.
For length scales smaller than some crossover, $r < l_{\times}$,
$C_q$ displays a non-trivial scaling behavior,
while at length scales larger than this crossover, $r > l_{\times}$,
$C_q$ is constant, $C_{q}(x)\approx C_{q}^{\rm sat}$.
In our previous work \cite{GSP}
it was shown that the simulated gel surfaces are multiaffine,
and the scaling exponents, crossover length scales, and saturation 
values of the increment correlation function 
are strongly dependent on the void concentration.
Despite the fact that the experimental surfaces seem to have a similar dependence on the void concentration,
one major difference between the simulated and experimental surfaces remains:
The experimentally measured surfaces are clearly self-affine on short 
length scales \cite{exp1,exp2},
while the simulated surfaces are clearly multiaffine \cite{GSP}.

Recent work \cite{Steven} indicates that vertical discontinuities
are one important source of multiaffinity, and the surfaces
generated by our model certainly have many vertical
discontinuities caused by overhangs. Even with a weakly
interacting AFM tip, experimental measurement of a soft surface,
such as the polyacrylamide gels studied in ~\cite{exp2}, should
have the effect of smoothing out any vertical discontinuities and
overhangs, thus showing a self-affine surface, even if the real 
surface is multiaffine. To mimic the effect of passing a weakly
interacting AFM tip over the spring-network surfaces, we have
smoothed the simulated surfaces by convolution with a Gaussian of
$\sigma=0.1$ in the dimensionless spatial units of the problem.
This process eliminates the overhangs without changing the overall
structural properties of the surfaces.

Figure~1 shows a detail of the generated surface and the smoothed surface.
Figure~2 shows the $q$th root of the generalized increment correlation function
$C_q^{1/q}(r)$ for systems with $0\%, 40\%$, and $49.5\%$ vacancies,
for both the original and smoothed surfaces.
It is evident that when the discontinuities are eliminated by smoothing,
the $H_q$ values become independent of $q$,
thus indicating that the surfaces are no longer multiaffine
but have become self-affine after smoothing.
The $H_q$ exponents are plotted in Fig.~3 versus the void concentration
for the original spring-network surfaces.
For the smoothed surfaces, $H_q = 1.0$ for all void concentrations,
as expected for a smooth surface.
The multiaffine scaling observed is consistent 
with that caused by vertical discontinuities \cite{Steven}.

\begin{figure}
\centerline{\includegraphics[width=0.5\columnwidth]{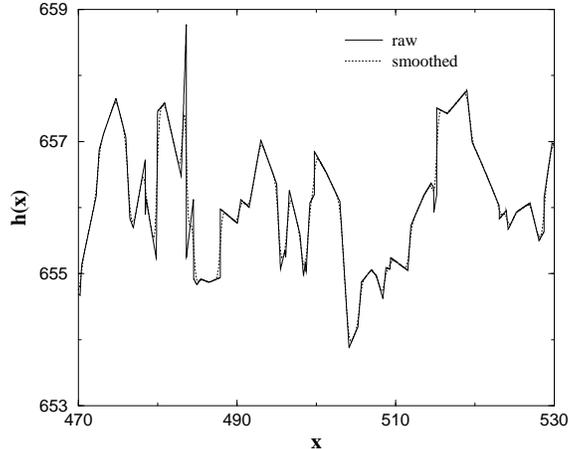}}
\caption[] {A typical portion of the spring-network surface (20\%
vacancies) before and after smoothing by convolution with a
Gaussian. The smoothing removes the vertical discontinuities
without significantly altering the overall surface features. Such
smoothing should mimic the measurement process of using a weakly
interacting AFM tip to probe the structure of a soft surface. }
\end{figure}

\begin{figure}
\centerline{\includegraphics[width=0.5\columnwidth]{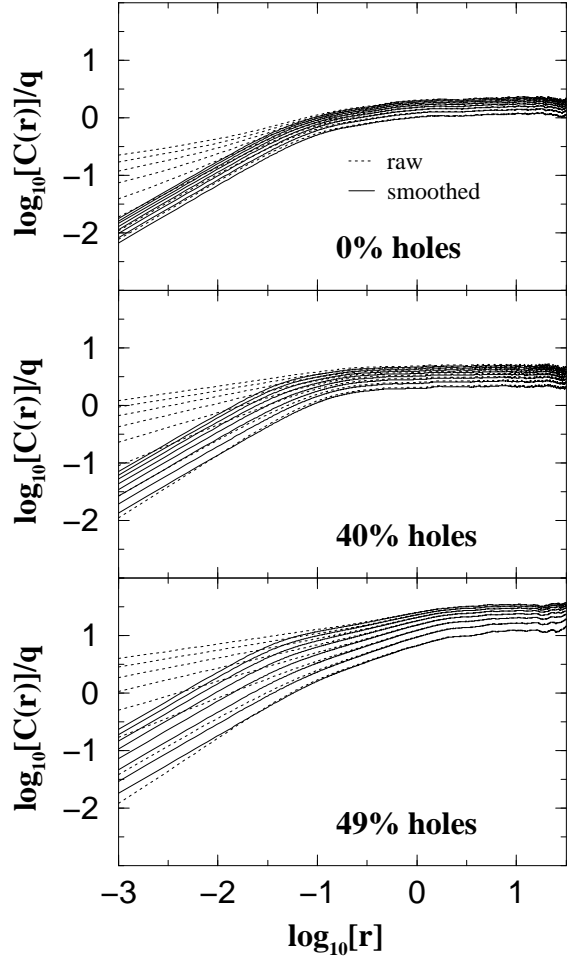}}
\caption[]
{
The $q$th root of $C_q(r)$ from the spring-network surfaces
for $q=0.5$ to $q=4.0$ in steps of 0.5.
For graphical simplicity, the $q$ labels have been omitted,
but $C_i(r)>C_j(r)$, when $i>j$.
The surfaces are averaged over 8, 8, 
and 10 independent realizations of the equilibrium spring lengths
for the networks with 0\%, 40\%, and 49\% vacancies, respectively.
The linear regions of the plot indicate power-law scaling,
and for the original simulated surfaces,
the different power-law exponents (slopes in the log-log plots)
indicate $q$-dependent multiaffine scaling.
After smoothing, the power-law scaling at small length scales is self-affine
(parallel lines in the log-log plots) with $H_q = 1$ for all $q$.
}
\end{figure}

\begin{figure}
\centerline{\includegraphics[width=0.5\columnwidth]{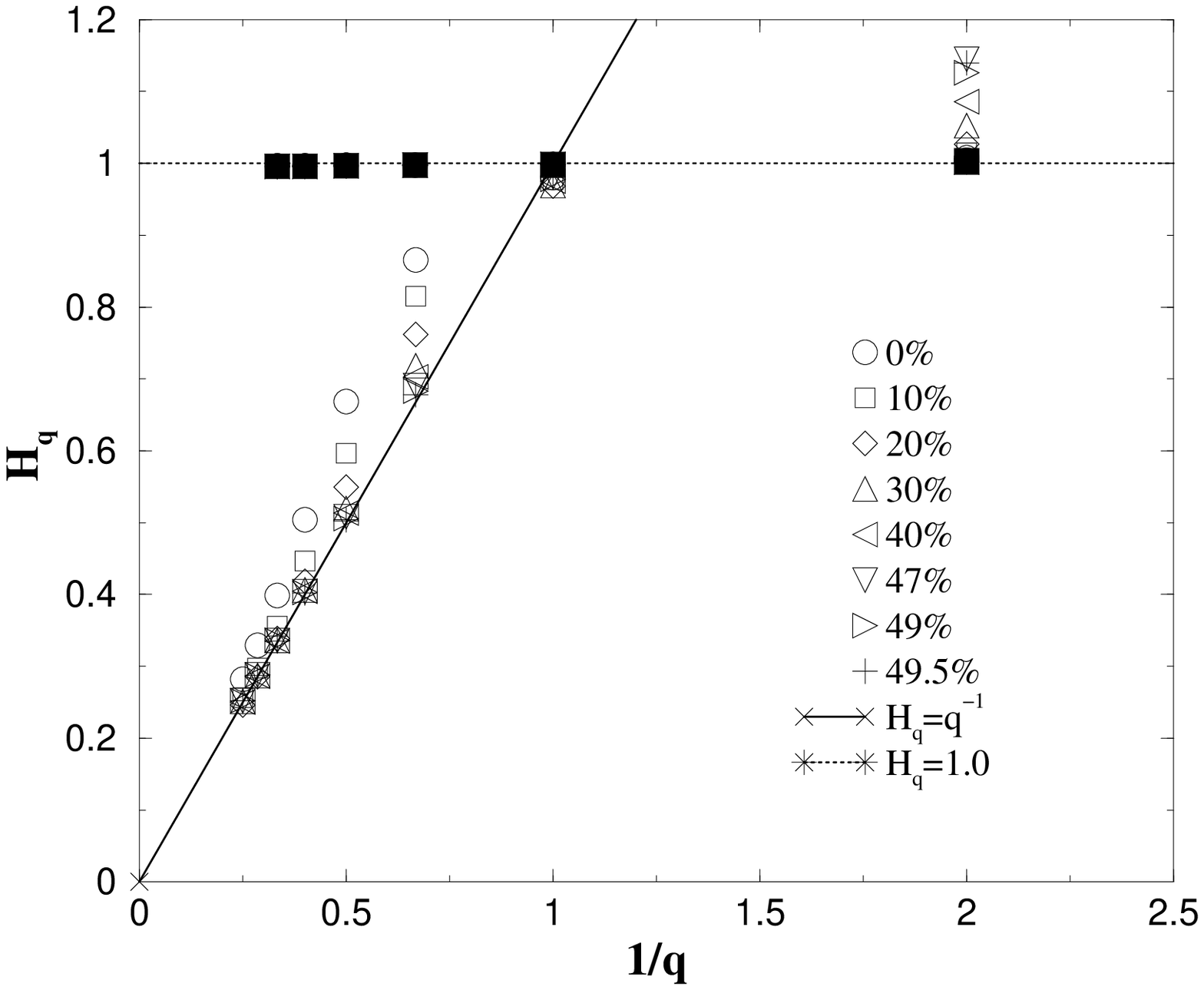}}
\caption[] { $H_q$ vs.\ $1/q$ for the spring-network surfaces. 
Open symbols indicate the original, multiaffine surfaces, while 
corresponding filled symbols indicate the smoothed, self-affine
surfaces. As expected, $H_q = 1.0$ for all of the smoothed surfaces. 
The observed multiaffine scaling of the original 
surfaces is consistent with the behavior caused by vertical
discontinuities as expected from Ref.~\cite{Steven}. The data are 
averaged over 8 independent realizations of the equilibrium spring
lengths for 0\% to 47\% vacancies, 10 realizations for 49\% vacancies, and
14 realizations for 49.5\% vacancies. }
\end{figure}

\section{Conclusions}

In this paper we have shown that surfaces generated by a
frustrated spring-network model can reproduce several aspects of
the scaling behavior of surfaces of surfactant-templated
polyacrylamide gels, as observed in recent AFM experiments
\cite{exp2}. Both real and simulated surfaces show a nontrivial
scaling behavior characterized by a power-law form of the
increment correlation function at small and intermediate length
scales. At larger length scales the increment correlation
function reaches a saturation value. This large-scale behavior
reflects the average network structure, while the small-scale 
behavior is due to microscopic density fluctuations.
Below some crossover length, the experimental surfaces are
self-affine while the original simulated surfaces are multiaffine.
We show that after smoothing by convolution with a narrow Gaussian, which
eliminates vertical discontinuities, the simulated surfaces become
self-affine. Such smoothing mimics the effect of passing a weakly
interacting AFM tip over the soft hydrogel surfaces. The large-scale
behavior is not affected by the smoothing process.




\end{document}